\documentclass[showpacs,preprint,aps]{revtex4}%
\usepackage{amsmath}
\usepackage{graphicx}
\usepackage{amsfonts}
\usepackage{amssymb}%
\begin{document}
\title{Heralded generation of symmetric and asymmetric entangled qudits with weak
cross-Kerr nonlinearity}
\author{Qing Lin}
\email{qlin@hqu.edu.cn}
\affiliation{College of Information Science and Engineering, Huaqiao University (Xiamen),
Xiamen 361021, China}


\begin{abstract}
High dimensional entangled states have attracted much more attentions, due to
their strong nonlocality and much powerful capability for quantum information
processing. By the methods presented in this paper, arbitrary forms entangled
qudits including symmetric and asymmetric forms could be generated with the
weak cross-Kerr nonlinearity. These schemes are heralded by the detections of
single-photon detectors. If all the detectors do not register any single
photons, the generation is success with the probability $1/n^{M}$ determined
by dimension $n$ and partite $M$. Furthermore, these schemes work well even
with the common photon number nonresolving detectors, therefore they are
feasible with the current experimental technology.

\end{abstract}
\maketitle

\section{Introduction}

Quantum entanglement acts as crucial resources in the area of quantum
information processing. It makes various quantum information tasks to be
possible, while these tasks, such as quantum teleportation \cite{tel}, quantum
dense coding \cite{coding}, etc., are impossible for classical information.
The generation of quantum entanglement will continue throughout the
development of quantum information processing. How to increase the length of
quantum entanglement and reinforce the ability to generate various forms of
quantum entanglement have been a long hotter topics in the research of quantum
entanglement generation. Recently, many exciting progress are reported to set
the new records of the length of quantum entanglement. In 2009, T. Monz
\textit{et al }reported the creation of 14-qubit entanglement with trapped
ions \cite{trapion}, and later in 2011, X. C. Yao \textit{et al} \cite{Pan}
and Y. F. Huang \textit{et al } \cite{Guo} had reported the generation of
8-photon entanglement, respectively. On the other hand, in the past years,
various forms of quantum entanglement have also been generated in optical
system, e.g., GHZ state \cite{GHZ1, GHZ2}, W state \cite{W1, W2}, cluster
state \cite{cluster}, etc. Besides of these traditional entangled qubits,
there is a special form of quantum entanglement, the so-called high
dimensional entangled states, or called entangled qudits. With the qudits and
entangled qudits, we could increase the security of quantum cryptography
\cite{cry1, cry2, cry3, cry4, cry5, cry6, cry7} and the efficiency of quantum
logic gates \cite{Toffoli1, Toffoli2}. Moreover, high dimensional universal
resources (e.g. AKLT states) for measurement based quantum computation can be
related to the ground states of certain many-body strongly correlated
Hamiltonian \cite{Gross, cai1, cai2, kal, cai3}. 

In this paper, we will focus on the optical system and consider how to
generate entangled qudit efficiently. There are many qudit definitions in
optical systems, what we concern is the polarization degree of freedom of
multi-photon qudit \cite{Polar1, Polar2, Polar3, Polar4, Polar5, Polar6,
Polar7, eqdit, bqtrit, Joo, biqutrit}. Using the photon number of vertical
polarization as encoding, the definition of polarization qudit could be
expressed as $\left\vert j\right\rangle _{n}\equiv\left\vert
(n-j-1)H,jV\right\rangle $, for $j=0,\cdots,n-1$, where $n$ is the dimension,
and H and V represent the horizontal and vertical polarizations. This
definition requires that the $n-1$ photons are in the same spatial and
temporal mode, therefore it is hard to operate each photon respectively, which
increases the difficulty to generate various forms of entangled qudits. If the
independent qudits are available, the success probability of entangled qudit
generation is only $1/n$ \cite{eqdit, xqdit}; while only the qubits and
entangled qubits are available, the success probability of entangled qutrit
generation is 3/16 in Ref. \cite{Joo} or 1/4 in Ref. \cite{xmqdit}. Moreover,
almost all the former entangled qudit generation schemes only involved the
case of symmetric entangled qudits, which could be expressed as follows,%
\begin{equation}%
{\displaystyle\sum\limits_{j=0}^{n-1}}
c_{j}|j\rangle_{n}\otimes|j\rangle_{n},
\end{equation}
where $%
{\displaystyle\sum\limits_{j=0}^{n-1}}
\left\vert c_{j}\right\vert ^{2}=1.$ However, the following asymmetric
entangled qudits are also valuable,%

\begin{equation}%
{\displaystyle\sum\limits_{j=0}^{n-1}}
c_{j}|j\rangle_{n}\otimes\left\vert (j+k)\operatorname{mod}n\right\rangle
_{n},
\end{equation}
where $k=1,\cdots,n-1$, and $%
{\displaystyle\sum\limits_{j=0}^{n-1}}
\left\vert c_{j}\right\vert ^{2}=1.$ Especially, when $c_{j}=\frac{1}{\sqrt
{n}}\tau^{j}$ (where $\tau=e^{i2\pi/n}$), the asymmetric entangled qudits and
the symmetric entangled qudits constitute the maximally entangled basis of two qudits.

Here we will use the cross-phase modulation (XPM) approach \cite{xpm1, xpm2,
xpm3} to develop the generation of asymmetric entangled qudits. Briefly, the
XPM approach bases on the interaction between a Fock state $\left\vert
n\right\rangle $ and a coherent state $\left\vert \alpha\right\rangle $,
resulting in the transformation $\left\vert n\right\rangle \left\vert
\alpha\right\rangle \rightarrow\left\vert n\right\rangle \left\vert \alpha
e^{in\theta}\right\rangle $, where the phase shift of coherent state is
determined by the photon number of the Fock state. Assisted with the XPM
approach, the generation scheme is heralded, and then the generated entangled
qudit could be used flexibly in the further quantum information processing. We
should note that our approach is available for the generation of symmetric
entangled qudits as well, and could be generalized to the case of
multi-partite, but not limit to only two-partite. 

The rest of the paper is organized as follows. In Section II, we first
introduce the ancilla single-photon qudit and then use the XPM approach to
generate asymmetric entangled qutrits. Then in the next section, we will
develop the approach to generate asymmetric entangled qudits. Sec. IV is for
discussion and conclusion remark.

\section{Generation of asymmetric entangled qutrits}

Before we outline the generation scheme, we will first introduce a
single-photon qudit, which is encoded by the spatial modes of the single
photon. After that, we will use the single-photon qudit as ancilla to generate
entangled qudits.

\subsection{Single-photon qudit}

\begin{figure}[ptb]
\includegraphics[width=12.7cm]{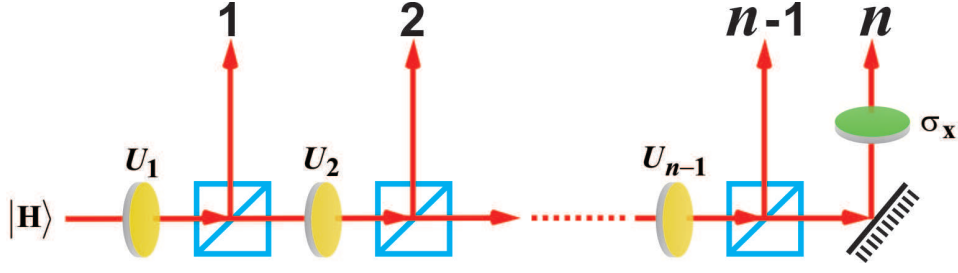}\caption{(Color online) Generation of
single-photon qudit. A single photon in the state $\left\vert H\right\rangle $
is injected into a series of proper unitary operations and polarized beam
spitters, the balanced single-photon qudit encoded by the spatial modes could
be generated.}%
\end{figure}

The balanced single-photon qudit could be expressed as follows,%

\begin{equation}
|\phi\rangle_{s}^{n}=\frac{1}{\sqrt{n}}%
{\displaystyle\sum\limits_{j=0}^{n-1}}
|j\rangle_{s},
\end{equation}
where $|j\rangle_{s}$ denotes the spatial mode of the single-photon. This
state could be generated easily by the setups shown in the Fig.1. The input
state is initially prepared as the state $\left\vert H\right\rangle $, and
then will be transformed into the state $\sqrt{\frac{n-1}{n}}\left\vert
H\right\rangle +\frac{1}{\sqrt{n}}\left\vert V\right\rangle $ by the first
single-photon unitary operation $U_{0}=\left(
\begin{array}
[c]{cc}%
\sqrt{\frac{n-1}{n}} & -\frac{1}{\sqrt{n}}\\
\frac{1}{\sqrt{n}} & \sqrt{\frac{n-1}{n}}%
\end{array}
\right)  $. After that, the state is injected into a polarized beam splitter
(PBS), which let the horizontal polarization $\left\vert H\right\rangle $ to
be passed and the vertical polarization $\left\vert V\right\rangle $ to be
reflected (denote this spatial mode to be $|0\rangle_{s}$). Repeating the
process, associated with the single-photon unitary operations $U_{j}=\left(
\begin{array}
[c]{cc}%
\sqrt{\frac{n-j-1}{n-j}} & -\frac{1}{\sqrt{n-j}}\\
\frac{1}{\sqrt{n-j}} & \sqrt{\frac{n-j-1}{n-j}}%
\end{array}
\right)  $ $\left(  j=1,\cdots,n-1\right)  $ and $\sigma_{x}$ (used in the
final spatial mode and then the polarization of all the spatial modes are the
same to be $\left\vert V\right\rangle $), the balanced single-photon qudit
could be generated.

\subsection{asymmetric entangled qutrit}

\begin{figure}[ptb]
\includegraphics[width=12.7cm]{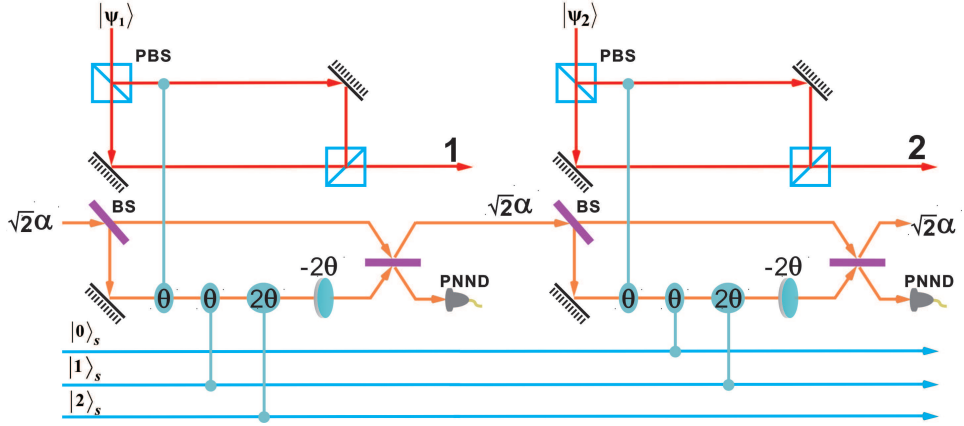}\caption{(Color online) Generation
of asymmetric entangled qutrit. Introduce a balanced single-photon qudit as
ancilla, which is coupled to two qubus beams through cross-phase modulation
(XPM) processes, associated with two independent polarized qutrits. With the
proper design of XPM phase shifts, the asymmetric entangled qutrits could be
heralded generated with the success probability 1/9. In addition, if more
independent qutrits are introduced and we repeat the same processes,
multi-partite entangled qutrits could be generated as well.}%
\end{figure}

To show our generation scheme clearly, we first use the generation of
asymmetric entangled qutrit as example and in follows will generalize it to
generate asymmetric entangled qudit. The balanced single-photon qutrit, as
well as two qubus beams $\left\vert \alpha\right\rangle \left\vert
\alpha\right\rangle $ are required as ancilla. The initial state is supposed
to be an independent bi-photon qutrit as $|\psi_{1}\rangle=a_{0}|0\rangle
_{3}+a_{1}|1\rangle_{3}+a_{2}|2\rangle_{3}$, where $%
{\displaystyle\sum\limits_{j=0}^{2}}
\left\vert a_{j}\right\vert ^{2}=1$. These independent qutrits could be
created by higher-order parametric down-conversion process \cite{bqtrit}, by
the Hong-Ou-Mandal (HOM) interference of two single-photon qubits \cite{HOM},
or by the transformation from single-photon qutrit \cite{xqutrit}. This qutrit
is injected into a PBS and the upper mode, associated with the modes
$|1\rangle_{s}$, $|2\rangle_{s}$ of the single-photon qutrit, will be coupled
to one of the qubus beam as depicted in Fig.2. The XPM phase shifts induced by
the couplings of the upper mode and the mode $|1\rangle_{s}$ are supposed to
be $\theta$, while that of the mode $|2\rangle_{s}$ is supposed to be
$2\theta$. After that, we will get the following state,%
\begin{align}
&  \frac{1}{\sqrt{3}}\left(  a_{0}|0\rangle_{3}|0\rangle_{s}+a_{1}%
|1\rangle_{3}|1\rangle_{s}+a_{2}|2\rangle_{3}|2\rangle_{s}\right)  \left\vert
\alpha\right\rangle \left\vert \alpha e^{i2\theta}\right\rangle \nonumber\\
&  +\frac{1}{\sqrt{3}}\left(  a_{0}|0\rangle_{3}|1\rangle_{s}+a_{1}%
|1\rangle_{3}|2\rangle_{s}\right)  \left\vert \alpha\right\rangle \left\vert
\alpha e^{i3\theta}\right\rangle +\frac{1}{\sqrt{3}}a_{0}|0\rangle
_{3}|2\rangle_{s}\left\vert \alpha\right\rangle \left\vert \alpha e^{i4\theta
}\right\rangle \nonumber\\
&  +\frac{1}{\sqrt{3}}\left(  a_{1}|1\rangle_{3}|0\rangle_{s}+a_{2}%
|2\rangle_{3}|1\rangle_{s}\right)  \left\vert \alpha\right\rangle \left\vert
\alpha e^{i\theta}\right\rangle +\frac{1}{\sqrt{3}}a_{2}|2\rangle_{3}%
|0\rangle_{s}\left\vert \alpha\right\rangle \left\vert \alpha\right\rangle .
\end{align}
Implement a displacement $-2\theta$ to the second qubus beam and let the two
qubus beams interfered on the 50:50 beam splitter (BS), which will implement
the transformation $\left\vert \alpha_{1}\right\rangle \left\vert \alpha
_{1}\right\rangle \rightarrow\left\vert \frac{\alpha_{1}-\alpha_{2}}{\sqrt{2}%
}\right\rangle \left\vert \frac{\alpha_{1}+\alpha_{2}}{\sqrt{2}}\right\rangle
$ will yield the following state,%
\begin{align}
&  \frac{1}{\sqrt{3}}\left(  a_{0}|0\rangle_{3}|0\rangle_{s}+a_{1}%
|1\rangle_{3}|1\rangle_{s}+a_{2}|2\rangle_{3}|2\rangle_{s}\right)  \left\vert
0\right\rangle \left\vert \sqrt{2}\alpha\right\rangle \nonumber\\
&  +\frac{1}{\sqrt{3}}\left(  a_{0}|0\rangle_{3}|1\rangle_{s}+a_{1}%
|1\rangle_{3}|2\rangle_{s}\right)  \left\vert \alpha_{-}^{1}\right\rangle
\left\vert \alpha_{+}^{1}\right\rangle +\frac{1}{\sqrt{3}}a_{0}|0\rangle
_{3}|2\rangle_{s}\left\vert \alpha_{-}^{2}\right\rangle \left\vert \alpha
_{+}^{2}\right\rangle \nonumber\\
&  +\frac{1}{\sqrt{3}}\left(  a_{1}|1\rangle_{3}|0\rangle_{s}+a_{2}%
|2\rangle_{3}|1\rangle_{s}\right)  \left\vert \alpha_{-}^{-1}\right\rangle
\left\vert \alpha_{+}^{-1}\right\rangle +\frac{1}{\sqrt{3}}a_{2}|2\rangle
_{3}|0\rangle_{s}\left\vert \alpha_{-}^{-2}\right\rangle \left\vert \alpha
_{+}^{-2}\right\rangle ,
\end{align}
where $\left\vert \alpha_{\pm}^{k}\right\rangle =\left\vert \frac{\alpha
\pm\alpha e^{ik\theta}}{\sqrt{2}}\right\rangle $ $\left(  k=\pm1,\pm2\right)
.$ If the vacuum component $\left\vert 0\right\rangle $ can be well
distinguished from the other components $\left\vert \alpha_{-}^{\pm1,\pm
2}\right\rangle $, we could achieve the following state,
\begin{equation}
a_{0}|0\rangle_{3}|0\rangle_{s}+a_{1}|1\rangle_{3}|1\rangle_{s}+a_{2}%
|2\rangle_{3}|2\rangle_{s}, \label{f1}%
\end{equation}
with the success probability 1/3. In this case, the qubus beams could be
recycled, since they are the same as the initial ones. The desired
discrimination could be realized by an idea photon number non-resolving
detector (PNND) (on/off detector with quantum efficiency $\eta=1$) and the
corresponding error probability due to the overlap of the vacuum component and
the other components is
\begin{equation}
P_{E}=\frac{4}{9}e^{-2\left\vert \alpha\right\vert ^{2}\sin^{2}\frac{\theta
}{2}}+\frac{2}{9}e^{-2\left\vert \alpha\right\vert ^{2}\sin^{2}\theta},
\label{p2}%
\end{equation}
which could tend to $0$ under the conditions $\left\vert \alpha\right\vert
^{2}\sin^{2}\frac{\theta}{2}\gg1$ and $\left\vert \alpha\right\vert ^{2}%
\sin^{2}\theta\gg1$. This requirement could be satisfied by increasing the
amplitude of coherent state $\alpha$, when it works in the weak nonlinearity
regime ($\theta\ll1$).

Evidently, if we introduce another independent bi-photon qutrit and repeat the
above processes, we could get the symmetric entangled qutrit. On the other
hand, if we want to obtain the asymmetric entangled qutrit, the experimental
setups must be adjusted slightly, which is shown in the right side of Fig.2.
Firstly, we introduce another bi-photon independent qutrit which is prepared
as $|\psi_{2}\rangle=b_{0}|0\rangle_{3}+b_{1}|1\rangle_{3}+b_{2}|2\rangle_{3}%
$, where $%
{\displaystyle\sum\limits_{j=0}^{2}}
\left\vert b_{j}\right\vert ^{2}=1$. Similarly, this qutrit is injected into a
PBS and the upper mode, associated with the modes $|0\rangle_{s}$,
$|1\rangle_{s}$ of the ancilla single-photon qutrit, is coupled to the second
qubus beam as depicted in Fig.2. After that, the following state could be
achieved,%
\begin{equation}
\left(  a_{0}b_{1}|0\rangle_{3}|1\rangle_{3}|0\rangle_{s}+a_{1}b_{2}%
|1\rangle_{3}|2\rangle_{3}|1\rangle_{s}+a_{2}b_{0}|2\rangle_{3}|0\rangle
_{3}|2\rangle_{s}\right)  \left\vert \alpha\right\rangle \left\vert \alpha
e^{i2\theta}\right\rangle +rest.,
\end{equation}
where $rest.$ denote the other components that the qubus beam catches
different phase shifts other than $2\theta$. One more displacement $-2\theta$
is implemented and the interference of two qubus beams will yield the state,%
\begin{equation}
\left(  a_{0}b_{1}|0\rangle_{3}|1\rangle_{3}|0\rangle_{s}+a_{1}b_{2}%
|1\rangle_{3}|2\rangle_{3}|1\rangle_{s}+a_{2}b_{0}|2\rangle_{3}|0\rangle
_{3}|2\rangle_{s}\right)  \left\vert 0\right\rangle \left\vert \sqrt{2}%
\alpha\right\rangle +rest.
\end{equation}
Obviously, using an ideal PNND to detect the first coherent state component,
the above state could be projected into the following state,%
\begin{equation}
a_{0}b_{1}|0\rangle_{3}|1\rangle_{3}|0\rangle_{s}+a_{1}b_{2}|1\rangle
_{3}|2\rangle_{3}|1\rangle_{s}+a_{2}b_{0}|2\rangle_{3}|0\rangle_{3}%
|2\rangle_{s},
\end{equation}
with the success probability $\left\vert a_{0}b_{1}\right\vert ^{2}+\left\vert
a_{1}b_{2}\right\vert ^{2}+\left\vert a_{2}b_{0}\right\vert ^{2}$. If the
coefficients $a_{i}=b_{i}=1/\sqrt{3}$, for $i=0,1,2,$ the success probability
is also 1/3. Moreover, it is easy to find that the corresponding error
probability is the same as Eq. (\ref{p2}).

Finally, to achieve the desired asymmetric entangled qutrit, we should erase
the ancilla single-photon qutrit without changing anything else. We firstly
perform the following Fourier transformation by a so-called linear optical
multi-port interferometer (LOMI) \cite{LOMI} on the single-photon qutrit,
\begin{equation}
\left\vert j\right\rangle _{s}=\frac{1}{\sqrt{3}}%
{\displaystyle\sum\limits_{k=0}^{2}}
e^{2\pi ijk/3}\left\vert k\right\rangle _{s},
\end{equation}
where $j,k$ denote the spatial modes. After performing some phase shifts
controlled by the detection on this single photon through the classical
feedforward, we could get the desired asymmetric entangled qutrit,
\begin{equation}
a_{0}b_{1}|0\rangle_{3}|1\rangle_{3}+a_{1}b_{2}|1\rangle_{3}|2\rangle
_{3}+a_{2}b_{0}|2\rangle_{3}|0\rangle_{3}.
\end{equation}

By properly setting the coefficients $a_{i},b_{i}$, the following maximally
entangled qutrits could be obtained,%
\begin{equation}
\frac{1}{\sqrt{3}}\left(  |0\rangle_{3}|1\rangle_{3}+\tau^{m}|1\rangle
_{3}|2\rangle_{3}+\tau^{2m}|2\rangle_{3}|0\rangle_{3}\right)  ,
\end{equation}
where $\tau=e^{2\pi i/3}$ and $m=0,1,2$. Totally, the success probability is
1/9. Moreover, exchanging the order of two qutrits and properly setting the
coefficients, we could achieve the following maximally entangled qutrits,%
\begin{equation}
\frac{1}{\sqrt{3}}\left(  |0\rangle_{3}|2\rangle_{3}+\tau^{m}|1\rangle
_{3}|0\rangle_{3}+\tau^{2m}|2\rangle_{3}|1\rangle_{3}\right)  .
\end{equation}
These six asymmetric entangled qutrits associated with the other three
symmetric entangled qutrits constitute the maximally entangled basis of two
qutrits. We should note here that the generation scheme does not require any
postselection processes, that is it is heralded by the detection of the
coherent state component. If the detector registers any signals, the
generation scheme is failure, while no detection means the success of the
generation. Therefore, the generated entangled qutrit could be used flexibly
in the further quantum information processing.

\section{GENERATION OF ENTANGLED QUDITS}

\begin{figure}[ptb]
\includegraphics[width=12.7cm]{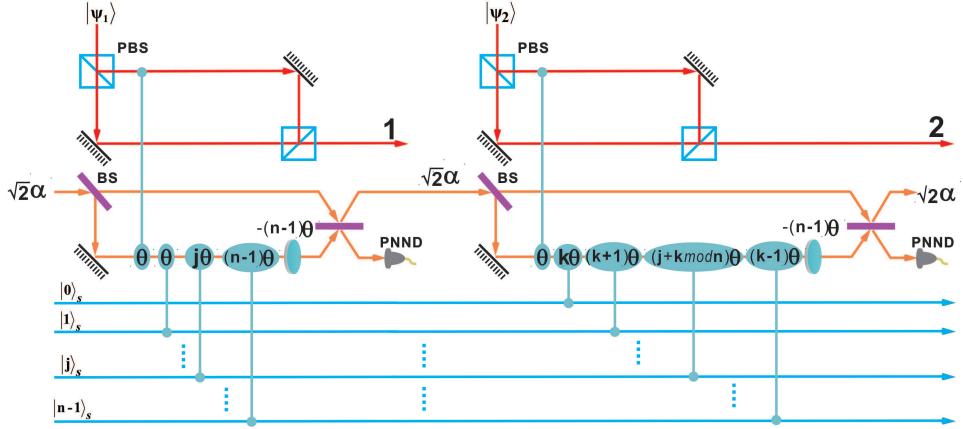}\caption{(Color online) Generation of
asymmetric entangled qudit. Similar to the generation of asymmetric entangled
qutrits, a balanced single-photon qudit is introduced as anciila. Two
independent polarized qudits are coupled to the two qubus beams, associated
with the ancilla single-photon qudit as depicted in Fig. 3. By the proper
design of XPM phase shifts, the asymmetric entangled qudits could be heralded
generated with the success probability $1/n^{2}$ determined by the dimension
$n$. Here we suppose $k\geq1$, which corresponds to the case of asymmetric
entangled qudits generation. If the XPM phase shifts induced to the two
independent qudits are the same, the symmetric entangle qudits could be
generated. Moreover, if more independent qudits are introduced and the similar
processes are repeated, multi-partite entangled qudits could be generated with
the success probability $1/n^{M}$ determined by the dimension $n$ and partite
$M$.}%
\end{figure}

The above scheme could be generalized to realize the generation of arbitrary
forms of entangled qudits (expressed in Eq. (2)). At first, we introduce a
single-photon qudit $|\phi\rangle_{s}^{n}$ as ancilla. Suppose an independent
multi-photon qudit is prepared as $%
{\displaystyle\sum\limits_{j=0}^{n-1}}
a_{j}\left\vert j\right\rangle _{n}$, where $%
{\displaystyle\sum\limits_{j=0}^{n-1}}
\left\vert a_{j}\right\vert ^{2}=1$. After the multi-photon qudit is injected
into a PBS, its upper mode associated with the spatial modes of the
single-photon qudit are coupled to the second qubus beam as depicted in Fig.4.
The XPM phase shift induced by the upper mode is supposed to be $\theta$ and
those by the spatial modes $\left\vert j\right\rangle _{s}$ ($j=0,\cdots,n-1$)
are supposed to be $j\theta$. After the interaction, we will get the following
state,%
\begin{equation}
\frac{1}{\sqrt{n}}%
{\displaystyle\sum\limits_{j=0}^{n-1}}
a_{j}|j\rangle_{n}|j\rangle_{s}\left\vert \alpha\right\rangle \left\vert
\alpha e^{i\left(  n-1\right)  \theta}\right\rangle +rest.,
\end{equation}
where $rest.$ denote the components that the qubus beam catches the phase
shifts other than $\left(  n-1\right)  \theta$. After that, implement a
displacement $-\left(  n-1\right)  \theta$ to the second qubus beam and let
the two qubus beams interfered on a 50:50 BS. The above state will be
transformed into the follows,%
\begin{equation}
\frac{1}{\sqrt{n}}%
{\displaystyle\sum\limits_{j=0}^{n-1}}
a_{j}|j\rangle_{n}|j\rangle_{s}\left\vert 0\right\rangle \left\vert \sqrt
{2}\alpha\right\rangle +rest.
\end{equation}
Using an ideal PNND to detect the first coherent state component will project
the above state to
\begin{equation}%
{\displaystyle\sum\limits_{j=0}^{n-1}}
a_{j}|j\rangle_{n}|j\rangle_{s},
\end{equation}
when the detection is $p_{n}=0$. The corresponding success probability is
$1/n$.

Similarly, we should introduce another independent multi-photon qudit, which
is supposed to be $%
{\displaystyle\sum\limits_{j=0}^{n-1}}
b_{j}\left\vert j\right\rangle _{n}$. If we want to achieve the symmetric
entangled qudit, what we should do is to repeat the above processes with the
same experimental setups. While we want to achieve the asymmetric entangled
qudit, we should adjust the experimental setups. As shown in the right side of
Fig.3, the corresponding XPM phase shifts induced by the upper mode of second
multi-photon qudit is also supposed to be $\theta$, while the other XPM phase
shifts induced by the spatial modes of the single-photon qudit are supposed to
be $\left[  \left(  j+k\right)  \operatorname{mod}n\right]  \theta$. After the
interaction, we could get the following state,%
\begin{equation}%
{\displaystyle\sum\limits_{j=0}^{n-1}}
a_{j}b_{(j+k)\operatorname{mod}n}|j\rangle_{n}\left\vert
(j+k)\operatorname{mod}n\right\rangle _{n}|j\rangle_{s}\left\vert
\alpha\right\rangle \left\vert \alpha e^{i\left(  n-1\right)  \theta
}\right\rangle +rest.
\end{equation}
Next we implement a displacement $-\left(  n-1\right)  \theta$ to the second
qubus beam and let two qubus beams interfered on a 50:50 BS. Then the
following state could be achieved after detecting the first coherent state
component with the result $p_{n}=0$,%
\begin{equation}%
{\displaystyle\sum\limits_{j=0}^{n-1}}
a_{j}b_{(j+k)\operatorname{mod}n}|j\rangle_{n}\left\vert
(j+k)\operatorname{mod}n\right\rangle _{n}|j\rangle_{s}.
\end{equation}
Finally, the ancilla single-photon qudit must be erased before we get the
desire quantum state. To complete the erasure, we first implement the
following Fourier transformation,%
\begin{equation}
\left\vert j\right\rangle _{s}=\frac{1}{\sqrt{n}}%
{\displaystyle\sum\limits_{k=0}^{n-1}}
e^{2\pi ijk/n}\left\vert k\right\rangle _{s},
\end{equation}
by a $n$-port LOMI \cite{LOMI}. After that, detecting the single-photon qudit
and using the detection to controlled the conditional phase shifts through the
classical feedforward will yield the desired asymmetric entangled qudit,%
\begin{equation}%
{\displaystyle\sum\limits_{j=0}^{n-1}}
a_{j}b_{(j+k)\operatorname{mod}n}|j\rangle_{n}\left\vert
(j+k)\operatorname{mod}n\right\rangle _{n}.
\end{equation}
By properly setting the coefficients, we could get the following states,%

\begin{equation}
\frac{1}{\sqrt{n}}%
{\displaystyle\sum\limits_{j=0}^{n-1}}
\tau^{jm}|j\rangle_{n}\otimes\left\vert (j+k)\operatorname{mod}n\right\rangle
_{n},
\end{equation}
where $m,k=0,\cdots,n-1$ and $\tau=e^{i2\pi/n}$. These states constitute to be
the maximally entangled basis of two qudits and the corresponding success
probability is $1/n^{2}$.

Furthermore, if we introduce one more multi-photon qudit and repeat the above
processes, we could achieve three-partite entangled qudits. In other words,
the above scheme could be easily generalized to create any forms of
multi-partite entangled qudits, including symmetric and asymmetric
multi-partite entangled qudits. The corresponding success probability is
$1/n^{M}$, which determined by the dimension $n$ and partite $M$.

\section{Discussion and CONCLUSION}

In this paper, we propose a scheme to generate entangled qudits, including the
symmetric and asymmetric forms. Using a balanced single-photon qudit encoded
by its spatial modes as ancilla, the forms of generated entangled qudits could
be flexible by setting the proper XPM phase shifts. The success probability
for maximally entangled qudits is determined by the dimension and the partite
as $1/n^{M}$. It looks that to be not high. However, this scheme is heralded,
but not based on the post-selection. If any of the ideal PNNDs register any
signals, the scheme is failure; otherwise, it is successful. While many other
optical schemes are based on the post-selection, and then when we use the
generated quantum entanglement, we will encounter many limitations, e.g., no
further interference is allowed. In other words, only single-photon operations
are allowed to performed on the generated quantum entanglement. Obviously, the
generated entangled qudits in our scheme could be used without any limitations
in the further quantum information tasks, since the generation is heralded.

Actually, the probabilistic quantum gate is still valuable for quantum
information processing. For example, the universal resource, e.g., cluster
state could be generated efficiently even with the probabilistic quantum gate
\cite{Duan}. Developing the idea in Ref. \cite{Duan}, we could generate the
high dimension universal resource, e.g., AKLT states efficiently by the
probabilistic gate presented in this paper as well. 

Now, we discuss the feasibility of the present scheme briefly. The ideal PNND
used in this scheme seems unrealistic, however, we could use the common PNND,
e.g. silicon avalanche photodiode (APD), to replace the ideal PNND. Though the
quantum efficiency of common PNND is lower than 1 ($\eta_{APD}\sim0.7$), it
could works well in our scheme. What we want is to distinguish the vacuum
state from the other coherent components. If we choose the proper parameters,
e.g., the XPM phase shift $\theta=0.01$ and the amplitude of qubus beam
$\left\vert \alpha\right\vert =500$, the average photon number of the coherent
components $\left\vert \alpha_{-}^{\pm1}\right\rangle $ and $\left\vert
\alpha_{-}^{\pm2}\right\rangle $ are about 13 and 50, respectively. In this
case, the vacuum state could be well separated from the coherent components
even with the common PNND, though the corresponding error probability
$P_{E}\ll1$.

Finally, the core element of our scheme is the proper design of XPM phase
shifts based on the weak cross-Kerr nonlinearity. Since we had theoretically
demonstrated that a small XPM phase shift with high fidelity is feasible in
the weak nonlinearity regime \cite{xc}, even the so-called multi-mode effect
is considered, then the weak nonlinearity in the idealized single-mode picture
is valid. Associated the mature linear optical techniques, e.g. the
interference of two qubus beams, the detection of single photons, etc., our
scheme is feasible with the current experimental technology.

\begin{acknowledgments}
The author thank Ru-Bing Yang for helpful suggestions. This work was funded by
National Natural Science Foundation of China (Grant No.11005040), Program for
New Century Excellent Talents in Fujian Province University (Grant No.
2012FJ-NCET-ZR04), the Fundamental Research Funds for Central Universities
(Grant No. JB-SJ1007), the Key Discipline Construction Project of Huaqiao
University and the State Scholarship Fund from China Scholarship Council.
\end{acknowledgments}

\end{document}